\documentclass[a4paper]{article}

\usepackage{graphicx}

\newcommand{\be}{\begin{equation}}
\newcommand{\ee}{\end{equation}}

\newcommand{\beq}{\begin{equation}}
\newcommand{\eeq}{\end{equation}}    
\newcommand{\calA}{\mathcal{A}}

\begin{document}

\begin{flushright}
ZTF--99/01
\end{flushright}

\begin{center}\Large \bf
LONG VS. SHORT DISTANCE DISPERSIVE TWO-PHOTON $K_L\rightarrow \mu^+ \mu^-$ 
AMPLITUDE\footnote{Presented by K. K. on the conference \emph{``Nuclear
and Particle Physics with CEBAF''}, Dubrovnik, Croatia, Nov 3-10, 1998}
\end{center}

\begin{center}
JAN O. EEG$^a$, KRE\v{S}IMIR KUMERI\v{C}KI$^b$ and IVICA PICEK$^b$
\end{center}

\begin{center} \em
$^a$Department of Physics, University of Oslo, Norway\\
$^b$Department of Physics, Faculty of Science, University of Zagreb,
    Croatia
\end{center}

\begin{center}
January 1999
\end{center}

% Abstract (it seems it should be formated just as a normal paragraph)
We report on the calculation of the two-loop electroweak, two-photon mediated
short-distance dispersive $K_L \rightarrow\mu^+\mu^-$ decay amplitude.
QCD corrections change the sign of this contribution and reduce it by
an order of magnitude. The resulting amplitude enables us to provide
a constraint on the otherwise uncertain long-distance dispersive amplitude.
% End of abstract

\vspace*{4mm}
\noindent
PACS numbers: 12.15.-y, 12.39.Fe \hfill   UDC: 539.12

\vspace*{1.5mm}
 
\vspace*{9mm}

The decay mode $K_L \to\mu^+\mu^-$ is a classical example of the rare
flavour changing neutral process that provided valuable insights into
the nature of weak interactions. Its non-observation at a rate
comparable with that of $K^{+}\rightarrow \mu^{+}\nu_{\mu}$ led to the
discovery of the GIM mechanism \cite{GIM} and to the derivation of the
early constraints on the masses of the charmed \cite{GL74} and, later,
top \cite{Be84} quark. 

Also, by studying this mode it could be possible to determine the
Wolfenstein $\rho$ parameter, to study the CP violation, and even to
discover some new physics (\emph{e. g.} through SUSY-induced FCNC
enhancement).  Because of this, this decay mode has received sustained
theoretical attention over the last three decades.

The lowest-order electroweak amplitude for $K_{L}\to \mu^{+}\mu^{-}$
in a free-quark  calculation \cite{GL74} (Fig.~1a and Fig.~1b) is
represented by one-loop (1L) W-box and Z-exchange diagrams,
respectively, and exhibits  a strong GIM cancellation.  Therefore, one
is addressed to consider the two-loop (2L) diagrams with photons in
the intermediate state (Fig.~1c) as a potentially important
contribution.

\begin{figure}[h]
\centerline{\includegraphics[width=10.0cm]{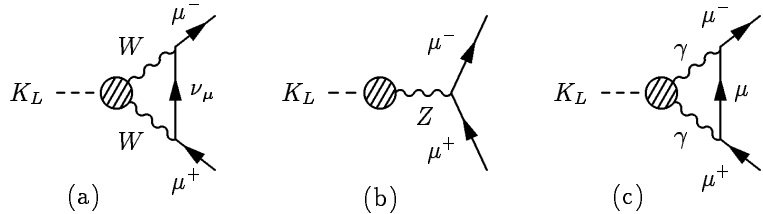}}
\caption{\em Possible mechanisms for $K_{L}\rightarrow\mu^{+}\mu^{-}$.}
\end{figure}
 
\noindent
If we normalize the amplitude $\calA$ to the branching ratio:
\be
B(K_L \to\mu^+\mu^-)=|\mbox{Re} {\cal A}|^2 + |\mbox{Im} {\cal A}|^2
\;,
\label{bri}
\ee
then the absorptive (Im$\calA$) part, which is dominated by the
process $K_L \to\gamma\gamma \to\mu^+\mu^- $ (Fig.~1c) with the
\emph{real} photons, is easily calculable and gives the so called
unitarity bound \cite{unit}
\be
B(K_L \to\mu^+\mu^-) \ge |\textrm{Im}\calA|^2=(7.1 \pm 0.2)\times 10^{-9} \; .
\label{babs}
\ee
corresponding to $|\mbox{Im} {\cal A}| = (8.4 \pm 0.1) \times 10^{-5}$.
If we compare this to the experimental number \cite{PDG}
\be
B(K_L \to\mu^+\mu^-)= (7.2 \pm 0.5)\times 10^{-9} \; ,
\label{branch}
\ee
we see that the  absorptive  part almost saturates the amplitude,
leaving only the small window for the dispersive (Re$\calA$) part
\begin{equation}
\mbox{Re}{\cal A} ={\cal A}_{\rm SD} + {\cal A}_{\rm LD}  \;,
\qquad\qquad |\textrm{Re}\calA|^2 < 5.6 \times 10^{-10} \;. 
\label{rebound}
\end{equation}
Thus, the total {\em real} part of the
amplitude, being the sum of short-distance (SD) and long-distance (LD)
dispersive contributions,
must be relatively small compared with the absorptive part of the
amplitude. Such a small total dispersive amplitude can be realized
either when the SD and LD parts are both small or by partial
cancellation between these two parts.

Now, the major obstacle in extracting useful short distance
information out of this decay mode is the poor knowledge of
$\calA_{LD}$.  There are several calculations of this LD part to be
found in literature \cite{BMS,KO,Va98,DAmIP98,GoDP98} and, later in
this paper, we will try to compare them. To this end it is necessary
to have a reliable estimate of the other, theoretically more
tractable, SD part $\calA_{SD}$.

Frequently, $\calA_{SD}$ has been identified as the weak contribution
represented by the one-loop W-box and Z-exchange diagrams of Figs. 1a
and 1b. This one-loop SD contribution ${\cal A}_{1L} = {\cal A}_{\rm
Fig.1a} + {\cal A}_{\rm Fig.1b}$  is dominated by the $t$-quark in the
loop (proportional to the small KM-factor $\lambda_{t}$), and the
inclusion of  QCD corrections  \cite{NSVZ,Buras} does not change this
amplitude essentially.  In the present paper we stress that the
diagrams of Fig.~1c, with \emph{virtual} intermediate photons,
with relatively high-momentum, lead to the same SD operator.
 That is, both the $1L$ diagrams contained in Figs. 1a and 1b,
as well as   $2L$ diagrams like those in Fig. 2, lead to the same
 SD operator of the type
\begin{equation}
  K_{SD} \;
  (\bar{d} \gamma^{\beta}Ls) (\bar{u}\gamma_{\beta}\gamma_5 v)  \;,
\label{2loop2}
\end{equation}
where $s,\bar{d},u,v$ are the spinors of the $s$- and $\bar{d}$
quarks in the $K$-meson, and the $\mu^+$ and $\mu^-$, respectively. 
The quantity $K_{SD}$ is a constant which contains the result of the 
SD calculations. The leading contributions from $2L$ diagrams are
$\sim \alpha_{\rm em}^2 G_{\rm F} \lambda_u$ and dominated by $c$-quarks
 in the loop, while the leading $1L$ is proportional to 
$G_{\rm F}^2 \, m_t^2 \, \lambda_t$.

 One should note that as already
pointed out in \cite{VS76,SV79}, the  two-loop diagrams
with two intermediate virtual photons have a short-distance part
${\cal A}_{\rm 2L}$ (contained in ${\cal A}_{\rm Fig.1c} = {\cal
A}_{\rm LD} + {\cal A}_{\rm 2L}$) that could pick up a potentially
sizable contribution, leading to the total SD amplitude is
${\cal A}_{\rm SD} ={\cal A}_{\rm 1L} +  {\cal A}_{\rm 2L}$.
By exploring the contribution from Fig.~1c leading to the 
${\cal A}_{\rm 2L}$ amplitude, we will be able to  isolate the 
strongly model-dependent LD dispersive piece.

\begin{figure}[ht]
\centerline{\includegraphics[width=10.0cm]{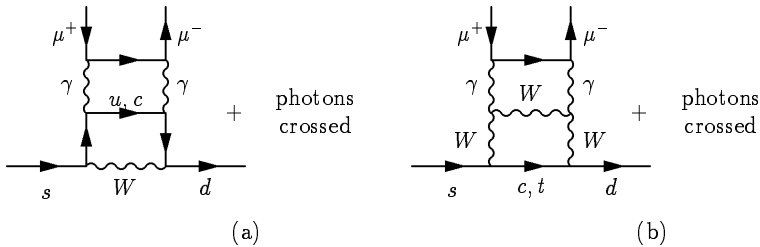}}
\caption{\em Typical two loop diagrams for $K_{L}\rightarrow\mu^{+}\mu^{-}$.}
\end{figure}

A complete treatment of the two-loop SD dispersive amplitude for
$K_{L}\to \mu^{+}\mu^{-}$ was given by us in Ref. \cite{EeKP98}.
There we used the momenta of the intermediate photons from the
diagrams in Fig.~1c to distinguish between SD and LD contributions, SD
part being defined by diagrams with photon momenta above some infrared
cut-off of the order of some hadronic scale $\Lambda\sim m_{\rho}$.
The fact that the resulting amplitudes depended only mildly on the
precise choice of $\Lambda$ assured us that the calculated amplitude
was sensible.

Our SD calculation in \cite{EeKP98} is dominated by the region 
$m_{\rho}< q^2 < m_{c}^2$ (the high
energy ($q^2 > m_{c}^2$) region is also included).
 After performing QCD corrections in the
leading logarithmic approximation \cite{EeKP98}, the original
electroweak amplitude was considerably suppressed and its sign changed:
\begin{equation}
   -0.38 \times 10^{-5} \leq {\cal A}_{\rm 2L}
  \leq  -0.001\times 10^{-5} \; ,
\label{result}
\end{equation}
where error bars stem mostly from empirical uncertainty
in $\alpha_{s}$.

Effectively, the LD calculation of the diagram on Fig.~1c is reduced
to the evaluation of the form-factor $F(q_{1}^{2}, q_{2}^{2})$
contained in the amplitude
\begin{equation}
A(K_{L}\to\gamma^{\ast}(q_1, \epsilon_1)\gamma^{\ast}(q_2, \epsilon_2))
= i \varepsilon_{\mu\nu\rho\sigma}\epsilon_{1}^{\mu}\epsilon_{2}^{\nu}
q_{1}^{\rho}q_{2}^{\sigma} F(q_{1}^{2}, q_{2}^{2}) \;,
\end{equation}
where $q_{1}^{2},q_{2}^{2}\neq 0$ measure the virtuality of the
intermediate photons.

The low energy regime $q^2 < \Lambda^2 \sim m_{\rho}^2$ is explorable
by chiral techniques determining $F(0,0)$.
In the standard $SU(3)_L \otimes SU(3)_{R}$ ChPT, where $\eta'$
is absent, one recovers the cancellation owing to the Gell-Mann-Okubo
mass relation, $\sim (3M_{\eta}^2 + M_{\pi}^2
- 4 M_{K}^2 ) \to 0 $. Keeping the $\eta'$ pole contribution in the
enlarged $U(3)_L \otimes U(3)_{R}$ symmetric theory \cite{GoDP98}
there is a destructive interference between the $\eta$ and
$\eta'$ contributions, so that the final amplitude is dominated
by the pion pole.

If going beyond the ChPT, one faces model calculations, and in
particular the calculations based on vector meson dominance (Refs.
\cite{BMS,KO} for example).The chiral-quark model may also be
used for the LD regime. Some preliminary analysis within the
chiral quark model indicates that the dispersive LD amplitude 
is of the same order of magnitude as the SD.

Combining Eqs. (\ref{rebound}) and (\ref{result}),   and 
$\calA_{1L}$ \cite{Buras,BuS98}, enables us to
find the following allowed range for ${\cal A}_{\rm LD}$:
\begin{equation}
   -0.1\times 10^{-5}  \leq {\cal A}_{\rm LD} \leq 6.5\times 10^{-5} \; \, .
\label{LDamp}
\end{equation}
Thus, having a  dispersive LD part 
${\cal A}_{\rm LD}$ of the size comparable with the  
absorptive part \cite{DAm86}
is still not ruled out completely.

The two vector-meson dominance calculations for the LD amplitude
 considered as the referent calculations in Ref. \cite{E791}
have basically opposite signs,  
\[
  -2.9\times 10^{-5} \leq {\cal A}_{\rm LD} \leq 0.5\times 10^{-5}
  \quad \cite{BMS}\;,
\]
\begin{displaymath}
   0.27\times 10^{-5} \leq {\cal A}_{\rm LD} \leq 4.7\times 10^{-5}
  \quad \cite{KO}\;,
\end{displaymath}
and the result of \cite{KO} seems to be more in agreement with the
bound (\ref{LDamp}). There are also some other, more recent, attempts
to calculate the ${\cal A}_{\rm LD}$ \cite{Va98,DAmIP98,GoDP98}.
The most stringent bound obtained is \cite{DAmIP98}
\begin{equation}
   |\textrm{Re}\calA_{LD}| < 2.9 \times 10^{-5} \; ,
\end{equation}
also well inside the allowed range (\ref{LDamp}).

\end{document}